\documentclass[aps,preprint,showpacs,floatfix]{revtex4-1}
\usepackage{bm}
\usepackage{amsmath}
\usepackage{epsfig}
\usepackage{rotating}
\usepackage{graphicx}

\def\pr{\prime}
\def\be{\begin{equation}}
\def\lan{\left\langle}
\def\ran{\right\rangle}
\def\ee{\end{equation}}
\def\barr{\begin{array}}
\def\earr{\end{array}}

\def\l{\left}
\def\r{\right}
\def\dis{\displaystyle}
\def\ed{\end{document}}

\def\cad{{\cal D}}

\def\dg{\dagger}

\def\ed{\end{document}}

\begin{document}

\title{Lie Algebraic approach to molecular spectroscopy: Diatomic to polyatomic
molecules}

\author{V.K.B. Kota\footnote{invited talk in the International Conference on 
Molecular Spectroscopy (ICMS 2017) held at Mahatma Gandhi University, Kottayam 
(Kerala, India) during 8-10 December, 2017 \\
Phone:+917926314939, Fax:+917926314460  \\
{\it E-mail address:} vkbkota@prl.res.in (V.K.B. Kota)}}

\affiliation{Physical Research Laboratory, Ahmedabad 380009, India}

\begin{abstract}

Interacting dipole ($p$) bosons along with scalar ($s$) bosons, based on the
ideas drawn from the interacting boson model of atomic nuclei, led to the
development of the vibron model based on $U(4)$ spectrum generating algebra for
diatomic molecules. The $U(4) \supset SO(4) \supset SO(3)$ algebra generates
rotation-vibration spectra. Extending this to two coupled $SO(4)$ algebras and
three $SO(4)$ algebras describe triatomic and four-atomic molecules
respectively. Similarly,  appropriately coupled $U(2) \supset SO(2)$ algebras
will describe the stretching vibrations, with proper point group symmetries, in
polyatomic  molecules. In addition, coupled $U(3)$ algebras describe coupled
benders. The Lie algebraic approach to molecular spectroscopy is briefly
described along with a list giving future directions and presented in three
appendices results for: (i) $U(3)$ algebra for bending vibrations and coupled
benders; (ii) symmetry mixing Hamiltonians generating regular spectra; (iii) 
partition functions for diatomic and triatomic molecules.

\end{abstract}


\maketitle
%\date{}

\section{Introduction}

Quantizing the relative co-ordinate in a diatomic molecules gives raise to a
description of vibrational-rotational spectra in terms of interacting dipole
$(\pi)$ bosons with $\ell=1^-$. The $\pi$ bosons along with scalar ($s$) bosons,
based on the ideas drawn from the interacting boson model of atomic nuclei
\cite{Iac-87,KS-17}, led to the development of the vibron model for diatomic
molecules with  $U(4)$ spectrum generating algebra (SGA) \cite{IL-95,FrV-94}.
The $SO(4)$ subalgebra  in $U(4) \supset SO(4) \supset SO(3)$ generates
rotation-vibration spectra; $SO(3)$ generates angular momentum. Extension with
two coupled $U(4) \supset SO(4)$ algebras describe stretching  and bending
vibrations in triatomic molecules.  Similarly, three coupled $U(4) \supset
SO(4)$ algebras describe four-atomic molecules. Continuing this to several
coupled $SO(4)$ algebras will in principle describe  polyatomic molecules but
these algebras will become unwieldy for molecules with 5 or more atoms. Then, an
alternative is to use coupled $U(2) \supset SO(2)$ algebras [it is also possible
to use the more complicated, but simpler than $U(4) \supset SO(4)$, coupled
$U(3)  \supset SO(3)$ algebras \cite{Iac-su3,Iac-su31,Iac-su32} as described in
Appendix-A].  This, along with a Majorana force will allow  for incorporating
the point group symmetries of polyatomic molecules within the Lie algebraic
approach and describe for example stretching vibrations in a variety of
polyatomic molecules \cite{IaOs-02,Rao-16}. Going beyond these and using the
ideas from the interacting boson-fermion model of atomic nuclei
\cite{IV-ibfm,KB,KD-lect}, Lie algebraic approach is also shown to describe
molecular electronic spectra \cite{Frank}. In this article we will give an
overview of these developments in the Lie algebraic approach to molecular
spectroscopy  with emphasis on group theoretical aspects. Now we will give a
preview.

Section 2 gives the results of $SO(4)$ algebra for diatomic molecules.
Similarly, Section 3 gives the results of coupled $SO(4)$ algebras for
triatomic and four-atomic molecules. Section 4 is on coupled $SU(2) \supset
SO(2)$ algebras for polyatomic molecules. Section 5 gives conclusions along with
a list giving future directions. These are supplemented with three appendices
describing the following in some detail: (i) $U(3)$ algebra for bending
vibrations and coupled benders and their application to quantum phase
transitions (QPT) and excited state quantum phase transitions (EQPT); (ii)
symmetry mixing Hamiltonians generating regular spectra using $U(4)$
subalgebras; (iii) partition functions for diatomic and triatomic molecules.

\section{SO(4) algebra for diatomic molecules}

Quantizing the relative co-ordinate $\vec{r}$ between the two atoms of a
diatomic molecule we have the vector boson ($\pi$ boson) with $\ell=1^-$;
$\pi^\dagger_\mu=(r_\mu -ip_\mu)/\sqrt{2}$ and $\pi_\mu=(r_\mu
+ip_\mu)/\sqrt{2}$. Now, introducing $s$ bosons ($\ell=0^+$) and demanding that
the total number ($N$) of $\pi$ and $s$ bosons is conserved, we have the vibron
model based with $U(4)$ spectrum generating algebra (SGA). The $U(4)$ is
generated by the 16 one-body operators $\pi^\dagger_\mu \pi_{\mu^\pr}$,
$s^\dagger s$, $\pi^\dagger_\mu s$, $s^\dagger \pi_{\mu^\pr}$. In angular
momentum coupled representation, introducing $\tilde{\pi}_\mu=(-1)^{1+\mu}
\pi_{-\mu}$ the number operator for $\pi$ bosons is $n_\pi=\sqrt{3} (\pi^\dagger
\tilde{\pi})^0$ and similarly, $n_s=s^\dagger s$. They will give the number of
$\pi$ bosons $N_\pi$ and $s$ bosons $N_s$ with $N=N_\pi + N_s$. The angular
momentum operator $L^1_\mu=\sqrt{2} (\pi^\dagger \tilde{\pi})^1_\mu$. Using the
commutation relations between the $U(4)$ generators, it is easy to see that
$U(4) \supset SO(4) \supset SO(3) \supset SO(2)$ where $SO(4)$ is generated by
the 6 operators  $L^1_\mu$ and $D^1_\mu=i(\pi^\dagger s +s^\dagger
\tilde{\pi})^1_\mu$, $SO(3)$ by $L^1_\mu$ and $SO(2)$ by $L^1_0$. Let us add
that it is also possible to have another $SO(4)$ algebra (called
$\overline{SO(4)}$) generated  by $L^1_\mu$ and $\cad^1_\mu= (\pi^\dagger s
-s^\dagger \tilde{\pi})^1_\mu$. We will not consider $\overline{SO(4)}$ any
further in this article except in Appendix B. The quantum numbers [called
irreducible representations (irreps) in the  representation theory of Lie
algebras] of $U(4)$, $SO(4)$ and $SO(3)$ are $N$, $\omega$ and $L$ respectively.
The $M$ quantum number of $SO(2)$ is trivial and it is dropped from now on as we
deal with only $L$ scalar Hamiltonians. The $N \rightarrow \omega \rightarrow L$
irrep reductions are easy to identify using pairing algebra in nuclear physics
and also using many other approaches \cite{IL-95,FrV-94,Nova}. Then we have, $N
\rightarrow \omega=N$, $N-2$, $N-4$, $\ldots$, $0$ or $1$ and $\omega
\rightarrow L=0$, $1$, $2$, $\ldots$, $\omega$. Using only the quadratic Casimir
invariants, the $U(4)$ Hamiltonian [assuming one plus two-body in nature and
preserving $N$ and $L$] for diatomic molecules ($H_{d-m}$) is
\be
\barr{rcl}
H_{d-m} & = & E_0 + \alpha C_2(SO(4)) + \beta C_2(SO(3)) \\
& = & E_0 + \alpha[L^2+D^2) + \beta L^2\;.
\earr \label{eq.1}
\ee
Here $E_0$ is a function of $N$. Using the well known formulas for the Casimir
invariants will give $E=E_0+\alpha \omega(\omega+2) + \beta L(L+1)$; note that
$\lan C_2(SO(4))\ran^{N,\omega,L} = \omega(\omega+2)$ \cite{KS-17}. Changing
$\omega$ into the vibrational quantum number $v=(N-\omega)/2$ will give the
energy formula to be
\be
\barr{l}
E=E_0^\prime-4\alpha (N+2)(v+\frac{1}{2})+4\alpha (v+\frac{1}{2})^2 + \beta
L(L+1)\;;\\
v=(N-\omega)/2 = 0,1,2,\ldots , \l[\frac{N}{2}\r]\;\mbox{or}\;
\l[\frac{N-1}{2}\r]\;,\\
v \rightarrow L=0,1,2,\ldots , (N-2v)\;.
\earr \label{eq.2}
\ee
Therefore, with $N$ large, $\alpha < 0$ and $\beta >0$, the $SO(4)$ algebra
generates rotation-vibration spectrum as seen clearly for example in H$_2$
molecule in its electronic ground state (here $N \sim 31$ and this follows from
the observed $v_{max}$ value). In fact $SO(4)$ represents rigid molecules (this
can be derived from the Morse oscillator) and the other limit $U(4) \supset 
[SU(3) \supset SO(3)] \oplus U(1)$ is for non-rigid molecules \cite{IL-95}; see
Appendix B. It is important to recognize that Eq. (\ref{eq.2}) is similar to the
well known Dunham expansion \cite{Dunham}.

\section{Coupled $SO(4)$ algebras for triatomic and four-atomic molecules}

Let us start with triatomic molecules. Now there are two relative co-ordinates
and associating $U(4)$ SGA to each of these, the SGA for triatomic molecules is
$U_1(4) \oplus U_2(4)$. This SGA admits large number of subalgebras but the most
important are: (i) local basis generated by $U_1(4) \oplus U_2(4) \supset
SO_1(4) \oplus SO_2(4) \supset SO_{12}(4) \supset SO(3)$; (ii) normal basis
generated by $U_1(4) \oplus U_2(4) \supset U_{12}(4) \supset SO_{12}(4) \supset
SO(3)$. In the local basis, the two $U_{i=1,2}(4)$ algebras give boson numbers
$N_1$ and $N_2$ and similarly the two $SO_{i=1,2}(4)$ give $\omega_1$ [or
$v_1=(N_1-\omega_1)/2$] from $N_1$ and $\omega_2$ [or $v_3=(N_2-\omega_2)/2$]
from $N_2$ respectively. The $SO_{12}(4)$ irreps are $(\tau_1,\tau_2)$ and they
are generated by the so called Kronecker product of $\omega_1$ and $\omega_2$.
This then gives (see for example \cite{Nova,IL-95} for the Kronecker products),
\be
\barr{l}
(\tau_1 , \tau_2) = \dis\sum_{\alpha , \beta} (\omega_1 + \omega_2 -\alpha
-\beta , \alpha - \beta)\;; \\
\alpha=0,1,\ldots , \mbox{min}(\omega_1,\omega_2),\;\;\beta = 0,1,\ldots,
\alpha\;.
\earr \label{eq.3}
\ee
Similarly, the reduction of $(\tau_1,\tau_2) \rightarrow L$ follows from the
recognition that $SO(4)$ is isomorphic to $SO(3) \otimes SO(3)$ and the two
$SO(3)$'s are labeled by $J_1=(\tau_1+\tau_2)/2$ and $J_2 = (\tau_1 -
\tau_2)/2$; $\tau_1 \geq \tau_2$. Then, the simple angular momentum
coupling rule gives $J_1 \times J_2 \rightarrow L$. The final result is
\be
\barr{l}
L= 0^+ , 1^- , 2^+ , \ldots, \tau_1^\pi\;;\;\;\mbox{for}\;\;\tau_2=0\;\;
\mbox{and}\;\;\pi=(-1)^{\tau_1} \\
L=\tau_2^\pm , (\tau_2+1)^\pm , \ldots , (\tau_1)^{\pm}\;;\;\;\mbox{for}\;\;
\tau_2 \neq 0\;.
\earr \label{eq.4}
\ee
More conventional notation for $(\tau_1,\tau_2)$ is to use $v_2^{\ell_2}$ with
$v_2=N_1+N_2-2v_1-2v_3-\tau_1$ and $\ell_2=\tau_2$. Using Eq. (\ref{eq.3}) we
have, $v_2=0,1,2,\ldots$, 2*min$(N_1-2v_1,N_2-2v_3)$ and $\ell_2=v_2, v_2-2,
\ldots$ $0$ or $1$. Note that $\ell_2=0,1,2,3,4,\ldots$ are in spectroscopic
notation $\Sigma$, $\Pi$, $\Delta$, $\Phi$, $\Gamma$ and so on. Adding the $L$ 
and $D$ operators from the two $SO(4)$'s will give the quadratic Casimir
invariant $L_{12}^2+D_{12}^2$ of $SO_{12}(4)$ and its eigenvalue  in
$(\tau_1,\tau_2)$ irrep are $[\tau_1(\tau_1+2) + \tau_2^2]$. Now, using
$H_{t-m}=E_0+a_1 C_2(SO_1(4)) + +a_2 C_2(SO_2(4)) +a_{12} C_2(SO_{12}(4)) + a_3
L_{12}^2$ will give a formula exactly similar to Dunham expression,
\be
E(v_1 v_2^{\ell_2} v_3 L)=E_0^\pr + \dis\sum_i \alpha_i (v_i+d_i)
+ \dis\sum_i \beta_i (v_i+d_i)^2 
+ \dis\sum_{i<j} \gamma_{ij} (v_i+d_i)
(v_j+d_j) + g_{12} l_2^2 + h L(L+1)\;.
\label{eq.5}
\ee
where $d_i=1/2$ for $v_1$ and $v_3$ and $1$ for $v_2$. For linear triatomic
molecules Eq. (\ref{eq.5}) is good. However, for bent molecules the projection
quantum number $k$ (same as $\ell_2$ but $\ell_2$ is used for linear molecules)
can take any value and different $k$ states are expected to be degenerate. Here,
we define $(v^{\prime}_2,k)$ via $\tau_1=N_1+N_2-2v_1-2v_3-2v^{\pr}_2-k$ and
$\tau_2=k$. Then, $v^{\pr}_2=0,1,2,\ldots$ and $k=0,1,2,3,\ldots$ for any
$v^\pr_2$. To obtain $k$ degeneracy, we need to consider
$\overline{C_2(SO_{12}(4))}=\sqrt{|L \cdot D|^2}$ and its eigenvalues in the
$(\tau_1 \tau_2)$ irrep are $\tau_2(\tau_1+1)$. Therefore, adding
$2a_{12}\overline{C_2(SO_{12}(4))}$ to $H_{t-m}$ will give $a_{12}
(\tau_1+\tau_2)(\tau_1+\tau_2+2)$ and then $E$ is independent of the $k$ quantum
number.

Turning to the normal mode basis, it is easy to identify that the $U_{12}(4)$
irreps will be $\{N_a,N_b\}=\{N_1+N_2-n,n\}$ where $n = 0, 1, 2, \ldots$,
min$(N_1,N_2)$. The $\{N_a,N_b\} \rightarrow (\tau_1 , \tau_2)$ reductions can
be written down but they involve the more complicated 'multiplicity' label; see
for example \cite{Nova,IL-95}. One usefulness of $U_{12}(4)$ is that it can be
used to mix local basis states and in reality, for describing linear or bent
molecules some mixing is essential. The Majorana interaction $M_{12}$, which is
related to $C_2(SU_{12}(4))$ in a simple manner and has a proper physical
meaning \cite{KS-17}, is added to $H$ for generating mixing. Inclusion of
$M_{12}$ term in $H_{t-m}$ is similar to Darling-Dennison coupling between the
local modes $v_1$ and $v_3$ \cite{IL-95}. In addition, also a Fermi coupling
term $F_{12}$ and higher order terms in Casimir operators are  added to $H$.
With these good agreements with data (within  1-5 cm$^{-1}$) are obtained for
many triatomic molecules such as H$_2$O, SO$_2$, CO$_2$, HCN, OCS, H$_2$S,
D$_2$O, N$_2$O and in some with different isotopes (ex: C$^{12}$O$_2$,
C$^{13}$O$_2$, H$_2$O$^{16}$, H$_2$O$^{18}$). Depending on the molecule,
$N_1=N_2$ or $N_1 \neq N_2$. Also, in all the cases the  value of $N_i$ is quite
large. Besides comparing spectra, the algebraic approach also allows for
calculating intensities of vibrational excitations; see
\cite{IL-95,Iac-tri,Sark-06} for details. All these extend to four-atomic
molecules as shown by Iachello {\it et al.}, by coupling three $SO(4)$ algebras,
in a series of papers analyzing for example  spectroscopic properties of the
molecules C$_2$H$_2$, C$_2$D$_2$,  C$_2$HD and HCCF \cite{Iac-tetra}. Note that,
here the coupling of the first two $SO(4)$ algebras will give $(\tau_1,\tau_2)$
irreps and then these are coupled to the $(\omega_3,0)$ irreps of the third
$SO(4)$ algebra. As $SO(4) \sim SO(3) \otimes SO(3)$, all the algebra here also
is carried out by exploiting angular momentum algebra.

\begin{figure}
\begin{center}
\includegraphics[width=0.4\linewidth]{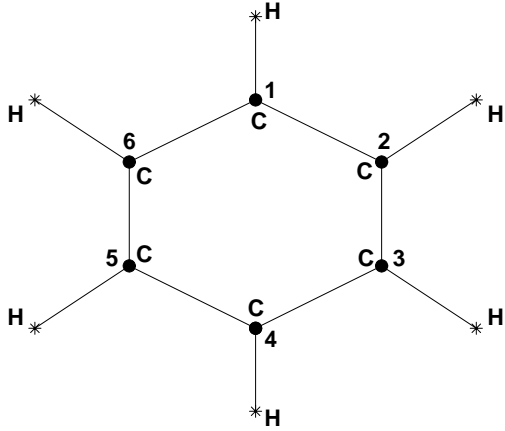}
\end{center}
\caption{Schematic figure showing Benzene molecule (C$_6$H$_6$) with $D_{6h}$
symmetry}
\label{fig1}
\end{figure}

\section{Coupled $U(2)$ algebras for vibrational modes in polyatomic molecules}

Study of the vibrational excited states in medium and large molecules is an
important current area of research. Based on the fact that $U(2) \supset SU(2)
\supset SO(2)$ [with boson number $N$ denoting $U(2)$ irreps, $\frac{N}{2}$ the
irreps of $SU(2)$ and $\frac{N}{2}-v$ the irreps of $SO(2)$] is the algebra of
one-dimensional Morse oscillator, a coupled $U(2)$ model for vibrational states
in polyatomic molecules has been introduced by Iachello and Oss \cite{IaOs-02}
by attaching a $U_i(2)$ algebra to each bond of a polyatomic molecule. Then, the
SGA for stretching vibrations is $\sum_i U_i(2) \oplus$. The interaction between
any two bonds $i$ and $j$ is then generated by (I) local $U_i(2) \oplus U_j(2)
\supset SO_i(2) \oplus SO_j(2) \supset SO_{ij}(2)$ algebra and (II) normal
$U_i(2) \oplus U_j(2) \supset U_{ij}(2) \supset SO_{ij}(2)$ algebra. Note that
for simplicity the $SU(2)$ is dropped everywhere but one need to remember that
$SU(2) \supset SO(2)$ algebra is the simple angular momentum algebra with the
$J$ quantum number being $\frac{N}{2}$ and the $J_z$ quantum number being
$m=\frac{N}{2}-v$. Then, the local basis is $\l.\l|N_i, v_i, N_j, v_j,\r.\ran$.
Each bond energy is generated by $C_i=[2J_z(i)]^2-N_i^2$ with
\be
\lan C_i \ran^{N_i,v_i}= -4(N_iv_i-v_i^2)\;.
\label{eq.8}
\ee
It is important to note that the one dimensional Morse oscillator is given by
$h_m=p^2/2\mu + D[1-\exp-ax]^2 = a_0 + a_1 C$. Similarly the pair energy
operator preserving (I) is $C_{ij} = [2J_z(i) + 2J_z(j)]^2-(N_i+N_j)^2$ and its
matrix elements are,
\be
\lan C_{ij} \ran^{N_i,v_i,N_j,v_j}= -4\l[(N_i + N_j)(v_i +v_j)-
(v_i +v_j)^2\r] \;.
\label{eq.9}
\ee
The interaction between the bonds $i$ and $j$ will mix the local (I) basis
states. A simple operator for this purpose is the Majorana operator $M_{ij}$
that is related to the Casimir operator of $SU_{ij}(2)$. The $M_{ij}$ operator
and its matrix elements (they will follow easily from the angular momentum
algebra),
\be
\barr{l}
M_{ij}=-\l\{2\l[J_z(i)J_z(j)-\frac{N_i}{2} \frac{N_j}{2}\r] + J_+(i) J_-(j) + 
J_-(i) J_+(j)\r\}\;,\\
\\
\lan N_i v_i N_j v_j \mid M_{ij} \mid N_i v_i N_j v_j \ran 
= (N_iv_j + N_jv_i -2v_iv_j)\;,\\
\\
\lan N_i v_i-1 N_j v_j+1 \mid M_{ij} \mid N_i v_i N_j v_j \ran 
= -\dis\sqrt{(N_j-v_j)(N_i-v_i+1)v_i(v_j+1)}\;,\\
\\
\lan N_i v_i+1 N_j v_j-1 \mid M_{ij} \mid N_i v_i N_j v_j \ran 
= -\dis\sqrt{(N_i-v_i)(N_j-v_j+1)v_j(v_i+1)}\;.
\earr \label{eq.10}
\ee
Now, diagonalizing the following Hamiltonian
\be
H=E_0+\dis\sum_i^n A_{i} C_i + \dis\sum_{i<j}^n A^\pr_{ij} C_{ij} +
\sum_{i<j}^n \lambda_{ij} M_{ij}
\label{eq.11}
\ee
in the local basis $\dis\prod_i\l.\l|N_i v_i\r.\ran$ will give the vibrational
energies. However, molecules carry point group symmetries (ex: octahedral $O_h$
for XY$_6$, $D_{6h}$ for C$_6$H$_6$) and they need to be incorporated in Eq.
(\ref{eq.11}). It is recognized that this can be done easily by imposing
restrictions on the parameters $A$, $A^\pr$ and more importantly on
$\lambda_{ij}$.

Let us consider the Benzene molecule C$_6$H$_6$ as shown in Fig. 1. There are
six bonds and they are all equal imposing the conditions $N_i=N$, $A_i=A$ and
$A^\pr_{ij}=A^\pr$. The $\sum_{i<j}^6 \lambda_{ij} M_{ij}$ term is constrained
by $D_{6h}$ symmetry depending on $(i,j)$ nearest neighbors, next nearest
neighbors and so on. Simplest choice is $S=\sum_{i<j}^6 \lambda_{ij} M_{ij}$
with $\lambda_{ij}=1$. Next is $S^\pr=\sum_{i<j}^6 \lambda_{ij} M_{ij}$ with
$\lambda_{ij}=1$ for nearest neighbors and zero otherwise. The nearest neighbors
are with $(ij)=(12)$, (16), (23), (34), (45) and (56). Third choice is  $S^{\pr
\pr}=\sum_{i<j}^6 \lambda_{ij} M_{ij}$ with $\lambda_{ij}=1$ for next nearest
neighbors and zero otherwise. The next nearest neighbors are with $(ij)=(13)$,
(15), (24), (26), (35) and (46). With these, the $H$ that generates states with
$D_{6h}$ symmetry is $H = E_0 + A C + A^\pr C^\pr + \lambda S + \lambda^\pr
S^\pr + \lambda^{\pr \pr} S^{\pr \pr}$ where $C=\sum C_i$ and $C^\pr =
\sum_{i<j} C_{ij}$. Instead of constructing the $H$ and diagonalizing it in the
local basis, it is also possible to directly construct the symmetry adopted
basis \cite{Chen}. The algebraic method is applied successfully to stretching
overtones of SF$_6$, WF$_6$ and UF$_6$ molecules, C-H stretching and C-H bending
vibration levels in C$_6$H$_6$ (also C$_6$D$_6$), CH stretches in $n$-alkane
molecules and so on; see \cite{IaOs-02,Rao-16,Ma-Os} and references therein.

\begin{figure}
\begin{center}
\includegraphics[width=0.45\linewidth]{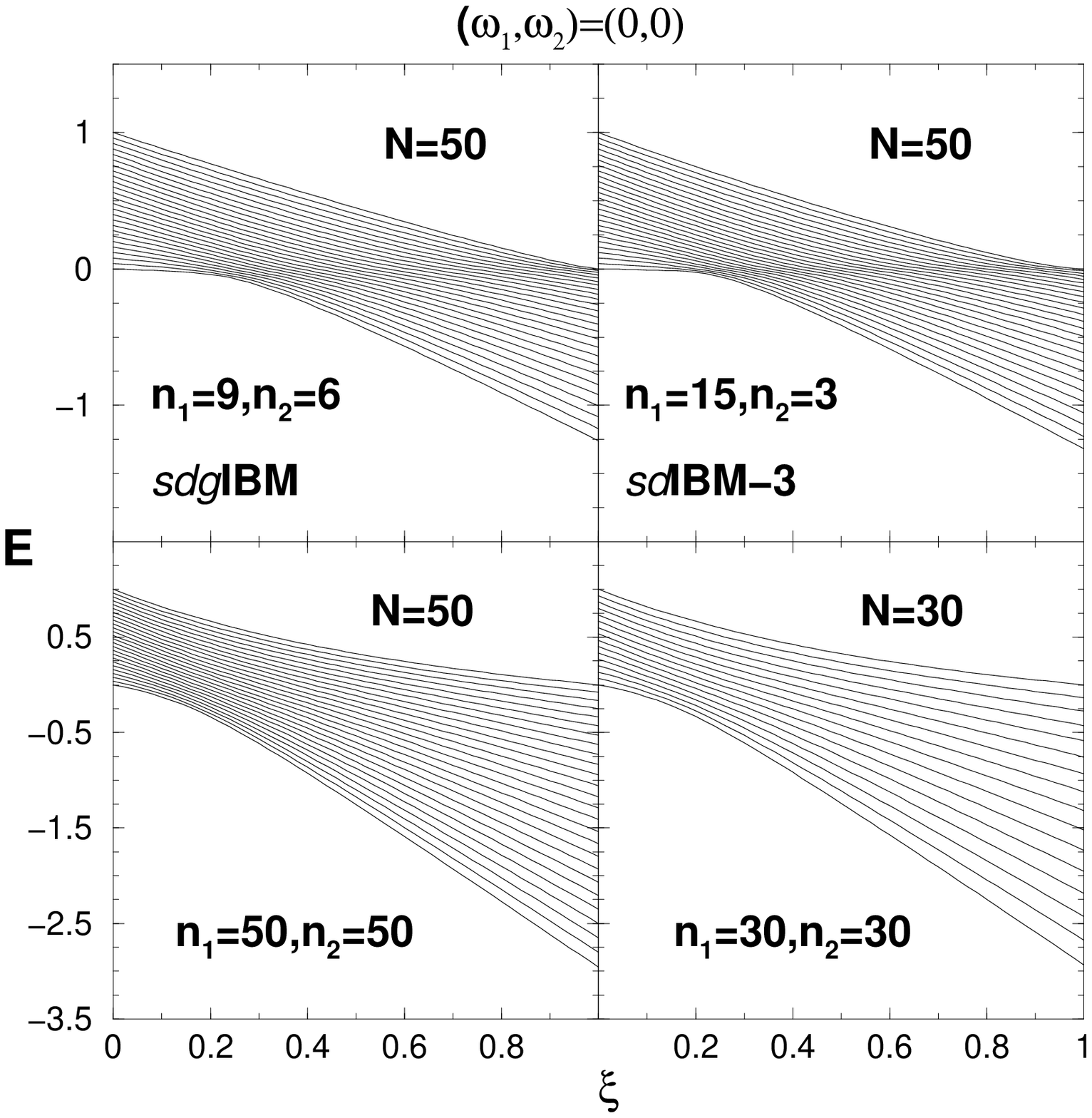} \\
\includegraphics[width=0.4\linewidth]{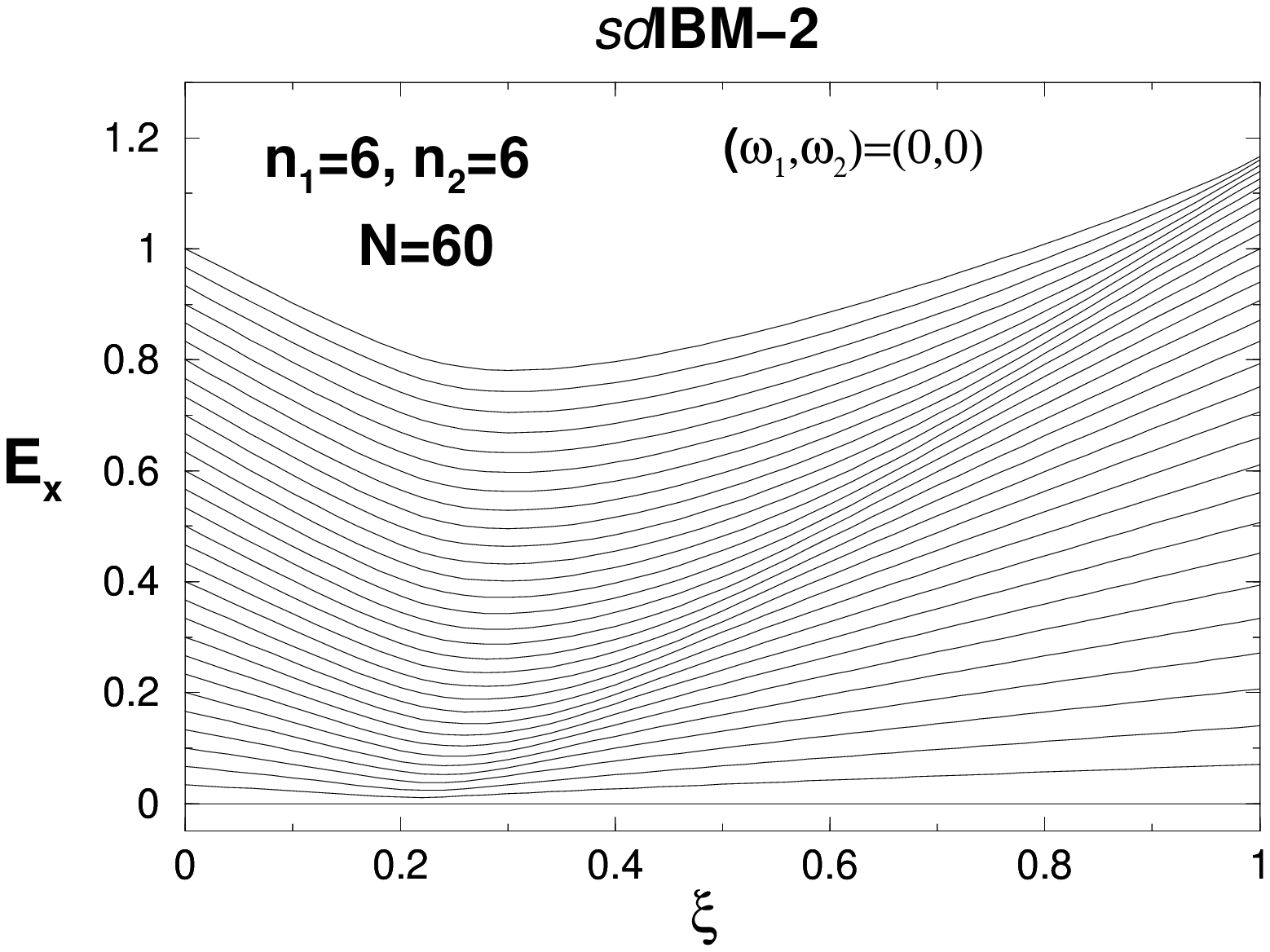} \\
\end{center}

\caption{(a) Spectra as a function of the mixing parameter $\xi$ in $H$
interpolating the symmetry limits (S1) and (S2) mentioned in Appendix A. Results
are shown for: (i)  $(n_1,n_2)$ = (9,6) and (15,3) with number of bosons N = 50;
(ii) $(n_1,n_2)$ = (50,50) with N = 50; (iii) $(n_1,n_2)$ = (30,30) with N = 30.
All results are shown for $(\omega_1, \omega_2) = (0,0)$ irrep where $\omega_1$
is the irrep of $SO(n_1)$ and $\omega_2$ is the irrep of $SO(n_2)$.  (b)
Excitation energies as a function of the mixing parameter $\xi$ for
$(n_1,n_2)$ =(6,6), N = 60 and $(\omega_1, \omega_2) = (0,0)$. Figures show that
there will be QPT only when the boson number $N$ is much greater than $n_1+n_2$.
Figures are taken from [28] and see this reference for further details.}

\label{fig2}
\end{figure}

\section{Conclusions and future outlook}

Starting with the $U(4) \supset SO(4) \supset SO(3)$ Lie algebra chain for
rotation-vibration levels in diatomic molecules, a brief account  of the Lie
algebraic approach to triatomic molecules using two coupled $SO(4)$ algebras,
for four-atomic molecules three coupled algebras and coupled $SU(2)$ algebras
for poly-atomic molecules is given in Sections II-IV. The Lie algebra approach
is not too complex and yet it is powerful as seen from the applications carried
out till now. In this short review, all mathematical details are kept to a
minimum and for detailed comparisons between theory and experimental data, the
reference given at the end should  be consulted. There are several new
directions enlarging the scope of the Lie algebra approach and some of these are
as follows. (i) Analysis of two coupled benders (in four-atomic molecules) using
coupled $U(3)$ algebras \cite{Iac-su3,Iac-su31,Iac-su32} and its extensions to
three or more benders. In two dimensions, necessary for describing bending
vibrations, $U(4)$ algebra reduces to $U(3)$ algebra and some details of the
$U(3)$ algebra are given in Appendix A. (ii) Algebraic approach for simultaneous
description of electronic, vibrational and rotational energy levels. For
example, with electrons in $s$ and $p$ orbitals the SGA for electrons is $U(8)
\supset U(4) \otimes SU(2)$ with $U(4)$ for the spatial part and $SU(2)$
generating spin. The key point now is that the spatial $U(4)$ can be combined
with the $U(4)$ generated by $(\pi,s)$ bosons to give a Bose-Fermi (BF) coupling
scheme \cite{Frank}. Let us add that BF schemes are well studied in nuclear
structure \cite{IV-ibfm,KB}. (iii) Development of the Lie algebra approach and
its applications to polyatomic molecules with very large number of atoms (also
to macromolecules, polymers etc.)  \cite{IaOs-02,Ma-Os,IaTr}. (iv) Shape phase
transitions that correspond to quantum phase transition (QPT) and excited state
quantum phase transitions (EQPT) can be studied using classical analysis of the
Lie algebraic models and with this it is possible to address quantum monodromy 
in molecules \cite{Monodr1,Iac-su31,Ca-RMP}. In fact quantum  monodromy is see
recently in some molecules \cite{Monodr2}. Some aspects of QPT and EQPT are
mentioned in Appendix A. (v) Applications of the algebraic coupling schemes 
discussed in Sections II-IV and Appendix-A in the study of order chaos
transitions and random matrix theory; see Appendix B and Ref. \cite{Ko-ee}. (vi)
Level densities, partition functions and other thermodynamic quantities for
polyatomic molecules can be studied using the algebraic models; some analytical
results for diatomic and triatomic molecules are available as presented in
Appendix C. 

\acknowledgments

Thanks are due to F. Iachello for discussions during many visits to Yale.

\begin{figure*}
\includegraphics[width=5in]{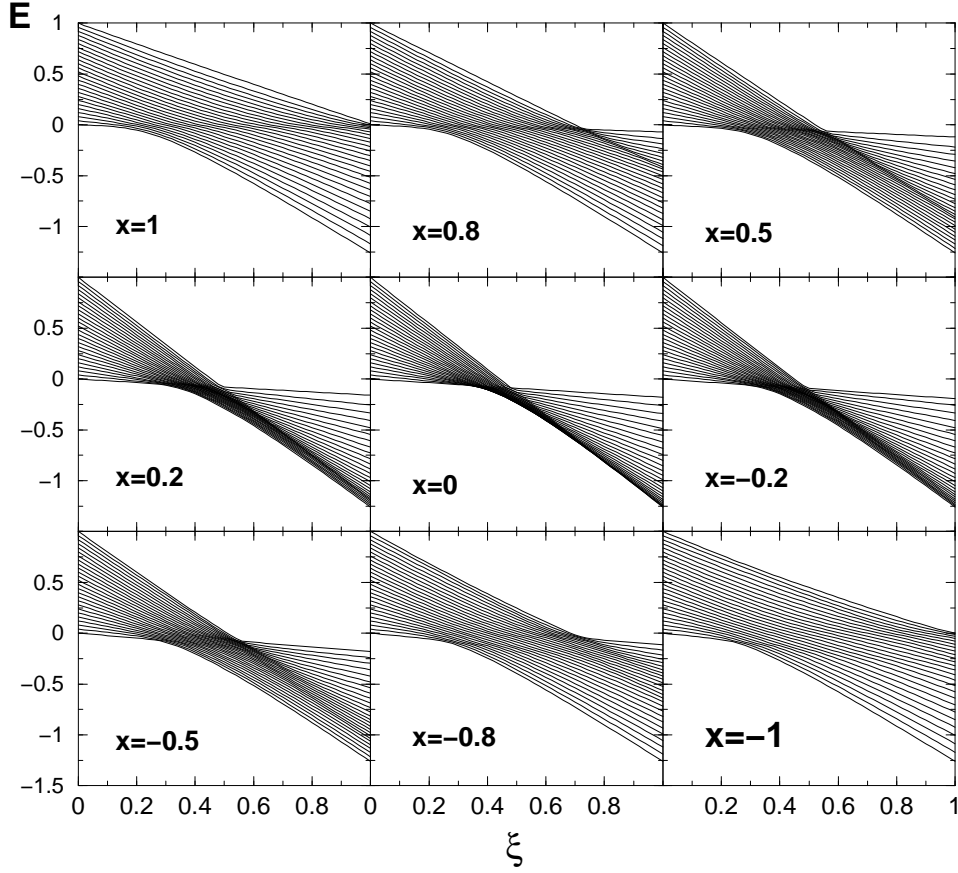}

\caption{Energy spectra for 50 bosons in $s$, $d$ and $g$ orbits with
$(\omega^B_{sd},\omega^B_g)=(0,0)$ in with the Hamiltonian $H_{sdg}(\xi ,x)=
[(1-\xi)/N^B]\,\hat{n}_g + [(\xi/(N^B)^2]\,[4(S_+^{sd}+xS_+^g)(S_-^{sd}+xS_-^g)
-N^B(N^B+13)]$ interpolating the symmetry limits (S1) and (S2) mentioned in
Appendix A with $n_1=6$ and $n_2=9$. Note that $S_+^{sd}$ is the pair creation
operator for the $sd$ boson system and $S_+^g$  for the $g$ boson system.
Similarly, $N^B$ is boson number operator, $\omega^B_{sd}$ is the $SO_{sd}$(6)
irrep and $\omega^B_g$ is the $SO_g(9)$ irrep. In each panel, energy spectra are
shown as a function of the parameter $\xi$ taking values from $0$ to $1$.
Results are shown in the figures for $x=1$, $0.8$, $0.5$, $0.2$, $0$, $-0.2$,
$-0.5$, $-0.8$ and $-1$; $x=1$ and $-1$ correspond to the two $SU(1,1)$ algebras
in the model. In the figures, energies are not in any units. Figure is taken
from [30].}  

\label{fig3}
\end{figure*}

\renewcommand{\theequation}{A-\arabic{equation}}
\setcounter{equation}{0}   

\begin{center} 
\section*{APPENDIX A}
$U(3)$ algebra chains for bending vibrations and coupled benders
\end{center}

In Sections II and III the full three dimensional $U(4)$ and coupled $U(4)$
algebras for diatomic to polyatomic molecules are briefly described and Section
IV coupled one dimensional $U(2)$ algebras for stretching vibrations are 
described. However,  even if one separates rotations and vibrations, one
dimensional description will not suffice for bending vibrations as these require
two dimensions, say $x$ and $y$. Introducing boson creation operators
$\tau^\dagger_x$ and $\tau^\dagger_y$ together with a scalar boson creation
operator $\sigma^\dagger$, we have three boson creation (call them
$b^\dagger_i$, $i=1$, 2 and 3 respectively) and three annihilation ($b_i$,
$i=1$, 2 and 3) operators. Then, clearly the SGA is $U(3)$ generated by the 9
operators $b^\dagger_i b_j$, $i,j=1,2,3$. In order to find the subalgebras in
$U(3)$, it is more convenient to consider circular bosons,
\be
\tau^\dagger_\pm = \dis\frac{1}{\dis\sqrt{2}} \l(\tau^\dagger_x \pm i
\tau^\dagger_y\r)\;,\;\;\; \tau_\pm = \dis\frac{1}{\dis\sqrt{2}} 
\l(\tau_x \mp i \tau_y\r)
\label{eq.ac1}
\ee
and they satisfy the commutation relations $[\tau^\dg_i\;\tau^\dg_j]=0$,
$[\tau_i\;\tau_j]=0$ and $[\tau_i\;\tau^\dg_j]=\delta_{ij}$. With these, the
number operator $\hat{n}$ giving number ($n$) of circular bosons is
\be
\hat{n} = \tau^\dg_x \tau_x + \tau^\dg_y \tau_y = \tau^\dg_+ \tau_+ + \tau^\dg_-
\tau_- = \hat{n}_+ +\hat{n}_- \;.
\label{eq.ac2}
\ee
Similarly, $\hat{n}_s = \sigma^\dg \sigma$ gives number of scalar bosons. The
total boson number $N=n+n_s$ is generated by $\hat{N}=\hat{n} + \hat{n}_s$.
Given $N$ bosons, it is easy to recognize that all the $N$ boson states belong 
to the totally symmetric irrep $\{N\}$ of $U(3)$. Also, It is well known that
$U(3)$ admits two subalgebras \cite{KD-lect}: (I) $U(3)  \supset SO(3) \supset
SO(2)$; (II) $U(3) \supset [U(2) \supset SU(2) \supset SO^\prime(2)] \otimes
U(1)$.  Let us now identify the generators of the various algebras in (I) and
(II) and the associated irrep reductions \cite{Iac-su3,Iac-su31}. 

Starting with (I), the $SO(3)$ algebra is generated by the three operators $D_+$,
$D_-$ and $D_0=\hat{\ell}$,
\be
D_+ = \dis\sqrt{2} \l(\tau^\dg_+ \sigma - \tau_-\sigma^\dg\r)\;,\;\;
D_- = \dis\sqrt{2} \l(\tau_+\sigma^\dg - \tau^\dg_-
\sigma\r)\;,\;\;D_0=\hat{\ell}=\hat{n}_+ - \hat{n}_-\;.
\label{eq.ac3}
\ee
This is established by proving easily that $[D_+\;,\;D_-]=2D_0$ and  $[D_0 \;,
\;D_+] = D_+$. The associated angular momentum quantum number is denoted by
$\omega$ and the eigenvalues of
\be
\hat{W}^2 = D_+ D_- +\hat{\ell}^2-\hat{\ell} = \frac{1}{2} \l(D_+D_- +
D_-D_+\r) + \hat{\ell}^2
\label{eq.ac4}
\ee
are $\omega(\omega+1)$. Then, the symmetry limit I is,
\be
\barr{l}
\l.\l| \barr{ccccc} U(3) & \supset & SO(3) & \supset & SO(2) \\
 \{N\} & & \omega  & & \ell \\
 & & \l(D_+ , D_- , \hat{\ell}\r) & & \hat{\ell}
\earr \r.\ran \\
\\
N \rightarrow \omega = N,N-2,\ldots,0\;\mbox{or}\;1 \\
\omega \rightarrow \ell=-\omega, -\omega+1,\ldots,0,\ldots,\omega-1,\omega\;.
\earr \label{eq.ac5}
\ee
As discussed ahead, $\omega$ quantum number is also related to pairing. As in
Section II, introducing the vibrational quantum number $v=(N-\omega)/2$ will
give $v=0,1,2,\ldots,(N/2)\;\mbox{or}\;(N-1)/2$ and $\ell=0, \pm 1, \pm 2,
\ldots, \pm (N-2v)$. Now, the basis states are $\l.\l|N,v,\ell\r.\ran$ and a
Hamiltonian (including at most quadratic Casimir invariants $C_r$, $r \leq 2$)
preserving the symmetry limit I is $H=E_0+\alpha C_1(U(3)) + A C_2(SO(3)) + B
[C_1(SO(2)]^2$ giving $E=E_0 + \alpha N +A N(N+1) -4A[(N+\frac{1}{2})v -v^2] + 
B \ell^2$. 

In $U(N) \supset SO(N)$ for bosons, the $SO(N)$ is related to pairing
\cite{Ko-00}. This result applies to $SO(3)$ in $U(3) \supset SO(3)$. With
$\tau_\pm$ and $\sigma$ bosons, the pair creation operator $\hat{P}$ is
\be
\hat{P} = 2\, \tau^\dg_+ \tau^\dg_- + \sigma^\dg \sigma^\dg \;.
\label{eq.ac5a}
\ee
Then, the pairing Hamiltonian, a two-body operator, is $H_p=\hat{P}
(\hat{P})^\dg$; note that $(\hat{P})^\dg = (2\tau_+ \tau_- + \sigma \sigma)$.
Simple algebra gives the important relation
\be
H_p = \hat{P} \l(\hat{P}\r)^\dg = \hat{N}\l(\hat{N}+1\r) - \l(\hat{W}\r)^2
\label{eq.ac5b}
\ee
establishing the relation between pairing and the $SO(3)$ algebra. It is also
important to point out that there is a second $SO(3)$ subalgebra in $U(3)$ and
we will denote this by $\overline{SO(3)}$. Its generators and the corresponding
pairing operator $H^\pr_P$ are
\be
\barr{l}
\overline{SO(3)}\,:\;(R_+ , R_- , \hat{\ell})\;, \\
R_+ = \dis\sqrt{2}\l(\tau^\dg_+\sigma + \tau_-\sigma^\dg\r)\;,\;\; 
R_- = \dis\sqrt{2}\l(\tau^\dg_-\sigma + \tau_+\sigma^\dg\r)\;,\;\;
\l[R_+\,,\,R_-\r]=2\hat{\ell}\;,\;\;\l[\hat{\ell}\,,\,R_+\r] = R_+\;,\\
\hat{R}^2 = R_+ R_- +(\hat{\ell})^2 -\hat{\ell} \rightarrow 
\lan \hat{R}^2 \ran^{N,\omega,\ell} = \omega(\omega+1)\;,\\
P^\pr=2\, \tau^\dg_+ \tau^\dg_- - \sigma^\dg \sigma^\dg\;,\;\;
H^\pr_P=P^\pr\l(P^\pr\r)^\dg = \hat{N}\l(\hat{N}+1\r) - \hat{R}^2\;.
\earr \label{eq.ac5c}
\ee
For the significance of $U(3) \supset \overline{SO(3)} \supset SO(2)$ see 
Appendix B.
   
Turning to limit II, it is easy to recognize that we can divide the space into
the one with $\tau$ bosons and other with $\sigma$ bosons giving $U(3) \supset
U_{\tau}(2) \oplus U_{\sigma}(1)$ with $U(2)$ generating $n$ and $U(1)$
generating $n_s$ so that $N=n+n_s$. As we always consider states with a fixed
$N$ value, given $n$ the  value of $n_s$ is uniquely  $N-n$ and therefore we
will not mention $U(1)$  hereafter. The $U(2)$ algebra is generated by the 4
operators
\be
Q_+ = \tau^\dg_+ \tau_-\;,\;\;Q_-=\tau^\dg_- \tau_+\;,\;\;Q_0=\frac{\hat{n}_+ -
\hat{n}_-}{2} = \frac{\hat{\ell}}{2}\;,\;\;\hat{n}\;.
\label{eq.ac6}
\ee
More importantly, the operators $\{Q_+ , Q_- , Q_0\}$ form angular momentum
algebra $SU(2)$ with $m$ quantum number $\ell/2$. It is easy to show that
$[Q_+\;,\;Q_-]=2Q_0$ and $[Q_0\;,\;Q_+]=Q_+$. Given $n$ bosons, the $SU(2)$ 
irrep is spin $\frac{n}{2}$. Then, 
\be
\dis\frac{\ell}{2} = -\dis\frac{n}{2}, -\dis\frac{n}{2}+1, \ldots,
\dis\frac{n}{2}-1, \dis\frac{n}{2} \Rightarrow \ell=\pm n, \pm (n-2), \ldots, 0
\;\; \mbox{or}\;\;1\;.
\label{eq.ac7}
\ee
Putting all these together, the symmetry limit II is,
\be
\barr{l}
\l.\l| \barr{ccccc} U(3) & \supset & U(2) & \supset & SO(2) \\
 \{N\} & & n  & & \ell \\
 & & \l(Q_+ , Q_- , \hat{\ell}/2, \hat{n}\r) & & \hat{\ell}
\earr \r.\ran \\
\\
N \rightarrow n = N,N-1,\ldots,0, \\
n \rightarrow \ell= \pm n, \pm (n-2), \ldots, 0\;\mbox{or}\;1\;.
\earr \label{eq.ac8}
\ee
Now, the basis states are $\l.\l|N,n,\ell\r.\ran$ and a Hamiltonian (including
at most quadratic Casimir invariants $C_r$, $r \leq 2$) preserving the symmetry
limit II is $H=E_0+\alpha C_1(U(2)) + \beta C_2(U(2))) + B [C_1(SO(2)]^2$ giving
$E=E_0 + \alpha n + \beta n(n+1) + B \ell^2$. 

Most general $U(3)$ Hamiltonian preserving $N$ and $\ell$ can be written as a
polynomial in the nine $U(3)$ generators $\hat{N}$, $\hat{n}$, $\hat{\ell}$,
$D_\pm$, $R_\pm$ and $Q_\pm$. Note that $D$ and $R$ operators change $\ell$ by
one unit and $Q$ by two units. It is easy to write the matrix elements of $H$
(i.e. construct $H$ matrix) in the $\l.\l|N,n,\ell\r.\ran$ basis; $n_s=N-n$,
$n_+=(n+\ell)/2$ and $n_-=(n-\ell)/2$. Both second degree and higher degree
polynomials are used in the applications to bending motion in many triatomic
molecules \cite{Iac-su31,Monodr1}. Another important aspect of the $U(3)$ model
is that the simple interpolating Hamiltonian
\be
H_{mix}=(1-\xi)\,\hat{n} + \dis\frac{\xi}{N-1} H_P 
\label{eq.ac9}
\ee
captures the essence of the two limits I and II. Note that $H_P$ is defined by 
Eqs. (\ref{eq.ac5a}) and (\ref{eq.ac5b}) and its matrix elements in the
$\l.\l|N,n,\ell\r.\ran$ basis follow easily from its definition,
\be
\barr{l}
\lan N, n^\pr , \ell \mid H_P \mid N,n,\ell\ran = \l[(N-n)(N-n-1) +n^2 
-\ell^2\r] \delta_{n^\pr , n} \\
+ \dis\sqrt{(N-n+2)(N-n+1)(n+\ell)(n-\ell)}\,\delta_{n^\pr , n-2} \\ 
+ \dis\sqrt{(N-n)(N-n-1)(n+\ell +2)(n-\ell +2)}\,\delta_{n^\pr , n+2}\;.
\earr \label{eq.ac10}
\ee
As shown in \cite{Iac-su31}, with $\xi$ varying from 0 to 1 the Hamiltonian
changes the structure from rigidly linear ($\xi=0$) to rigidly bent
($\xi=1$) structure. More importantly, for $0 < \xi \leq 2$, the molecule
will be quasi-linear and for $0.2 < \xi < 1$  quasi-bent. Moreover, at
$\xi=0.2$ the system exhibits QPT (change in ground state structure) and it is
a second order phase transition. Also, at $\xi=0.6$ the system with $H$
defined by Eq. (\ref{eq.ac9}) exhibits EQPT \cite{Monodr1}. Let us stress that
the $U(3)$ model is a simple two-level model ($3=2+1$) and the QPT and EQPT are
typical of general two level models \cite{Qpteqpt,Ca-RMP}. For bosons in two
levels with $n_1$ and $n_2$ number  of degenerate single particle levels, the
SGA is $U(n_1+n_2)$ and then there are two symmetry limits, S1: $U(n_1+n_2)
\supset U(n_1) \oplus U(n_2) \supset SO(n_1) \oplus SO(n_2) \supset K$ and S2:
$U(n_1+n_2) \supset SO(n_1+n_2) \supset SO(n_1) \oplus SO(n_2) \supset K$. In
generating the spectrum for a fixed $SO(n_1) \oplus SO(n_2)$ irrep, the Lie
algebra $K$ will not play any role. Numerical examples for QPT (also EQPT) are
shown in Figs. 2 and 3 for some general two level models. See
\cite{Ko-qpt1,Ko-qpt2,Ko-qpt3}  for details of the results in the figures.

Besides describing single benders, using coupled $U(3)$ algebras it is possible
to study various structures generated by coupled benders in tetra-atomic
molecules. Associating a $U(3)$ for each bender we have $U_1(3) \oplus U_2(3)$
SGA with large number of subalgebra chains preserving boson numbers $N_1$ and
$N_2$ and the total $\ell=\ell_1 + \ell_2$ quantum number. At the first level,
the subalgebras are $U_1(2) \oplus U_2(2)$, $SO_1(3) \oplus SO_2(3)$,
$U_{12}(3)$ and $U_1(2) \oplus SO_2(3)$. The $U_1(2) \oplus U_2(2)$ admits
$U_{12}(2)$ and $SO_1(2) \oplus SO_2(2)$ subalgebras, $SO_1(3) \oplus SO_2(3)$
admits $SO_{12}(3)$ and $SO_1(2) \oplus SO_2(2)$ subalgebra  [the later also
appears  in $U_1(2) \oplus SO_2(3)$] and finally $U_{12}(3)$ admits $U_{12}(2)$
and $SO_{12}(3)$ subalgebras. All these will have the final subalgebra
$SO_{12}(2)$. It is possible to write the generators of all these algebras and
also one and two-body operators that preserve $N$ and $\ell$. Extending the
algebras described before for one bender, it is possible to construct the $H$
matrix for the coupled benders systems. However, a simple Hamiltonian describing
the various structures is of the form $H=H_1 + H_2 + V_{12}$ with $H_i$ same as
discussed above for one bender and $V_{12}$ contains, $(P_{12})(P_{12})^\dg$,
$\hat{W_1} \cdot \hat{W_2}$, the quadratic Casimir invariant of $SU_{12}(3)$ or
equivalently the Majorana operator $M_{12}$ and so on. See \cite{Iac-su32,Iac-su33}
for further mathematical details and applications to C$_2$H$_2$ and H$_2$CO 
molecules.

\renewcommand{\theequation}{B-\arabic{equation}}
\setcounter{equation}{0}   

\begin{center} 
\section*{APPENDIX B}
Symmetry mixing Hamiltonians generating regular spectra
\end{center}

Given the two symmetry limits (i) $U(4) \supset SO(4) \supset SO(3)$ and (ii)
$U(4) \supset [SU(3) \supset SO(3)] \oplus U(1)$  for diatomic molecules, 
general two-body Hamiltonian mixing these two symmetry limits is,
\be
H_{mix}= \alpha_0(N) + \alpha_1 C_1(U(1)) + \alpha_2 C_2(SO(4)) + \alpha_3 
C_2(SU(3)) + \alpha_4 C_2(SO(3))\;.
\label{eq.b1}
\ee
Note that $\alpha_0(N)$ is a quadratic polynomial in $N$. More importantly, 
$\alpha_2=0$ will give limit (ii) and $\alpha_1=\alpha_3=0$ will give limit (i).
However, even when $\alpha_1,\alpha_2,\alpha_3 \ne 0$, it is possible to produce
a regular spectrum. This is  due to the existence of $\overline{SO(4)}$
generated by $L^1_q$ and $\cad^1_\mu= (\pi^\dagger s -s^\dagger
\tilde{\pi})^1_\mu$ mentioned in Section II. Note that both $C_2(SO(4))$ and
$C_2(\overline{SO(4)})$ generate the same spectrum with eigenvalues
$\omega(\omega+2)$. These operators are given by,
\be
\barr{l}
C_2(SO(4)) = 2\l(\pi^\dagger \tilde{\pi}\r)^1 \cdot \l(\pi^\dagger
\tilde{\pi}\r)^1 - (\pi^\dagger s +s^\dagger \tilde{\pi})^1 \cdot (\pi^\dagger 
s +s^\dagger \tilde{\pi})^1 \\
C_2(\overline{SO(4)}) = 2\l(\pi^\dagger \tilde{\pi}\r)^1 \cdot \l(\pi^\dagger
\tilde{\pi}\r)^1 + (\pi^\dagger s -s^\dagger \tilde{\pi})^1 \cdot (\pi^\dagger 
s -s^\dagger \tilde{\pi})^1 \\
\Rightarrow  C_2(SO(4)) + C_2(\overline{SO(4)}) = 4\l(\pi^\dagger \tilde{\pi}
\r)^1 \cdot \l(\pi^\dagger \tilde{\pi}\r)^1  -2\l[\pi^\dagger s \cdot s^\dagger
\tilde{\pi} + s^\dagger \tilde{\pi} \cdot \pi^\dagger s\r]\;.
\earr \label{eq.b2}
\ee
Using the results that $ss^\dagger=(n_s+1)$ and $\tilde{\pi} \cdot \pi^\dagger =
-(3+n_\pi)$ we have,
\be
C_2(\overline{SO(4)}) = -C_2(SO(4)) +4(N-1)n_\pi -4n^2_\pi +6N +2C_2(SO(3)) \;.
\label{eq.b3}
\ee
As $C_1(U(1))=n_s=N-n_\pi$ and $C_2(SU(3))=n_\pi (n_\pi +3)$, clearly for a
particular choice of the parameters in Eq. (\ref{eq.b1}), $H_{mix}$ can be
reduced to $C_2(\overline{SO(4)})$ and hence solvable (generates a regular
spectrum). For details of the significance of this result for order-chaos
transitions and quantum phase transitions, see \cite{Ca-RMP,Di-PRL}. It is also
important to add that the occurrence of multiple pairing algebras, as seen from
the $U(3)$ model discussed in Appendix A, is a general feature of both fermion
and boson systems with two or more levels or orbits and they play an important
role in QPT and EQPT; see \cite{Ko-qpt3}.

\renewcommand{\theequation}{C-\arabic{equation}}
\setcounter{equation}{0}   

\begin{center}
\section*{APPENDIX C} 
Partition functions for diatomic and triatomic molecules
\end{center}

Starting with the energy formula given by Eq. (\ref{eq.2}), it is possible to
derive a simple formula for the partition function $Z(\beta)=Tr(\exp -\beta E)$
for diatomic molecules in the $SO(4)$ [$U(4) \supset SO(4) \supset SO(3) \supset
SO(2)$] limit . Using Eq. (\ref{eq.2}) for the eigenvalues and the allowed
quantum numbers, we have
\be
Z_{SO(4)}(\beta) = Z_0\, \dis\sum_{v=0}^{[N/2]} \dis\sum_{L=0}^{N-2v}
(2L+1) \exp\,-\beta \{A(N+1)v - Av^2 +BL(L+1)\}\;.
\label{eq.a1}
\ee
Note that $A=-4\alpha$ and $B=\beta$; $\alpha$ and $\beta$ are defined in Eq.
(\ref{eq.2}). In addition, $Z_0$ is a constant. Clearly, with $A>0$ and $B>0$ 
the ground state is $\l.\l|N,v=0,L=0\r.\ran$. With $N \rightarrow \infty$ and
$\sigma=1/(2\beta B)^{1/2} >> 1$, $Z(\beta)$ takes the simpler form
\be
\barr{rcl}
Z_{SO(4)}(\beta) & \stackrel{N \rightarrow \infty , \sigma>>1}{\longrightarrow}
& Z_0\,\dis\sum_{v=0}^{\infty} \exp -\beta (AN)v \l\{\dis\int_{0}^{\infty} 
(2L+1) \exp -L(L+1)/2\sigma^2\,dL\r\} \\
& = & Z_0\,\l(2\sigma^2\r)\, \l(1-\exp-\beta (AN)\r)^{-1} \\
& = & Z_0\,Z_{rot}(\beta)\,Z_{vib}(\beta) \;. 
\earr \label{eq.a2}
\ee
Note that $Z_{rot}(\beta) = 2\sigma^2$, $Z_{vib}(\beta)=\l(1-\exp-\beta (AN)
\r)^{-1}$ and $\sigma^2=1/(2\beta B)$. The decomposition of $Z_{SO(4)}(\beta)$
into a product of $Z$'s for the rotational and vibrational parts is similar to
the decomposition  obtained before in the interacting boson model of atomic
nuclei \cite{Kota-90}. A different formula, in the limit $\beta \rightarrow 0$,
is given by
\be
\barr{l}
Z_{SO(4)}(\beta) \stackrel{\beta \rightarrow 0}{\longrightarrow} 
Z_0\,\dis\int_0^{N/2} dv\;\dis\int_{0}^{N-2v} dL (2L+1) \exp\,-\beta 
\{A(N+1)v - Av^2 +BL(L+1) \\
= Z_0\, \dis\frac{1}{B\beta}\dis\int_0^{N/2} dv \l\{1-\exp-\beta 
B(N-2v)(N-2v+1)\r\}\,\exp-\beta \l\{A(N+1)v -Av^2\r\}\;.
\earr \label{eq.a3}
\ee
The last integral here can be written in terms of error functions; see also 
\cite{Dimitri}. Let us add that more accurate formulas for $Z_{SO(4)}(\beta)$ 
can be derived using Euler-Maclaurin summation formula.

The other symmetry limit, for diatomic molecules, starting with $U(4)$ is  $U(4)
\supset [SU(3) \supset SO(3) \supset SO(2)] \oplus U(1)$ with  basis states
$\l.\l|N,n_\pi , L, M\r.\ran$ where $N \rightarrow n_\pi=0,1,\ldots,N$ and
$n_\pi \rightarrow L=n_\pi$, $n_\pi-2$, $\ldots$, $0$ or $1$. Now, the
Hamiltonian and the partition function in this $SU(3)$ limit are
\be
\barr{l}
H = E_0^{\pr \pr}+A_1 C_1(U(3)) + A_2 C_2(SU(3)) + A_3 L(L+1)\;,\\
Z_{SU(3)}(\beta) = Z_0\, \dis\sum_{n_\pi=0}^N \dis\sum_{L \in n_\pi} (2L+1)
\exp -\beta\l\{A_1 n_\pi + A_2 n_\pi (n_\pi +3) + A_3 L(L+1)\r\} \;.
\earr \label{eq.a4}
\ee 
Note that with $A_1>0$, $A_2 << A_3$ and $A_3 >0$, the ground state is $\l.\l| 
N, n_\pi =0, L=0\r.\ran$. In the symmetry limit it is a good approximation
to assume $A_2, A_3 \simeq 0$. Then we have,
\be
Z_{SU(3)}(\beta) = Z_0\,\dis\sum_{n_\pi=0}^N \dis\frac{(n_\pi +1)(n_\pi 
+2)}{2}\,\exp -\beta (A_1\,n_\pi) = Z_0\,(1-\exp-\beta A_1)^{-2}\;.
\label{eq.a5}
\ee
In addition, it is also possible to derive a formula for $Z_{SU(3)}(\beta)$ in
the $\beta \rightarrow 0$ limit in terms of error functions.

Turning to tri-atomic molecules, using the energy formula given by Eq.
(\ref{eq.5}) and the associated quantum numbers (see Section III), it is
possible to derive a formula for the partition function
$Z_{local-SO_{12}(4)}(\beta)$ in the local basis symmetry limit $U_1(4) \oplus
U_2(4) \supset SO_1(4) \oplus SO_2(4) \supset SO_{12}(4) \supset SO(3)$. In the
limit $N_1 \rightarrow \infty$, $N_2 \rightarrow \infty$, $a_{12} \sim 0$ and
$\sigma >> 1$ ($a_{12}$ is the strength of $C_2(SO_{12}(4))$ and
$\sigma^2=1/2\beta d$ where $d$ is the strength of $L(L+1)$ term), the energy
formula given by Eq. (\ref{eq.5}) reduces to the form $E(N_1, N_2, v_1,
v_2^{\ell_2}, v_3, L)=e_0+e_1v_1+e_2v_2+e_3v_3+dL(L+1)$. Then $Z(\beta)$ is,
\be
Z_{local-SO_{12}(4)}(\beta) = Z_0\,\dis\sum_{v_1,v_2,v_3=0}^{\infty} 
\dis\sum_{\ell_2 \in v_2} \dis\sum_L (2L+1)\, \exp-\beta\l\{e_1v_1+e_2v_2+
e_3v_3+dL(L+1)\r\}\;.
\label{eq.6}
\ee
The $L$ integration gives $2\sigma^2$ for $\ell_2=0$ and $2(2\sigma^2)$ for
$\ell_2 \neq 0$; see Eq. (\ref{eq.4}) for $\tau_2=\ell_2 \rightarrow L$ and the
doubling for $\ell_2 \neq 0$. Combining this with the $v_2 \rightarrow \ell_2$
reductions ($v_2=0 \rightarrow \ell_2=0$,  $v_2=1 \rightarrow \ell_2=1$, $v_2=2
\rightarrow \ell_2=0,2$, $v_2=3 \rightarrow \ell_2=1,3$, $v_2=4 \rightarrow
\ell_2=0,2,4$,  $\ldots$)   will allow us to carry out the $\ell_2$ summation in
Eq. (\ref{eq.6}) giving,
\be
\barr{rcl}
Z_{local-SO_{12}(4)}(\beta) & = & (2\sigma^2) \dis\sum_{v_1,v_2,v_3=0}^{\infty}
(v_2+1)\, \exp-\beta(e_1v_1+e_2v_2+e3v_3) \\
& = & (2\sigma^2)\l(1-\exp -\beta e_1\r)^{-1}\,\l(1-\exp -\beta e_3\r)^{-1}\,
\l(1-\exp -\beta e_2\r)^{-2}\;.
\earr \label{eq.7}
\ee
Further improvements of the formula for $Z_{local-SO_{12}(4)}(\beta)$ are
possible. Also, in future it is important to derive the formulas for $Z(\beta)$
for the other symmetry limits of the $U_1(4) \oplus U_2(4)$ model.

\ed